\documentclass[twocolumn,amsmath,superscriptaddress,prl,amssymb,aps,floatfix]{revtex4-1}

\usepackage{graphicx}
\usepackage{booktabs}
\usepackage{multirow}
\DeclareMathOperator{\Tr}{Tr}

\begin{document}

\title{Quantum density anomaly in optically trapped ultracold gases}
\author{Eduardo O. Rizzatti}
\affiliation{Instituto de F\'isica, Universidade Federal do 
Rio Grande do Sul, Porto Alegre-RS, Brazil}
\author{Marco Aur\'elio A. Barbosa}
\affiliation{Programa de P\'os-Graduac\~ao em Ci\^encia de 
Materiais, Universidade de 
Bras\'ilia, Planaltina-DF, Brazil}
\author{Marcia C. Barbosa}
\affiliation{Instituto de F\'isica, Universidade Federal do 
Rio Grande do Sul, Porto Alegre-RS, Brazil}

\date{\today}

\begin{abstract}
\indent \textit{We show that the Bose-Hubbard Model exhibits an increase in density with temperature at fixed pressure in the regular fluid regime and in the superfluid phase. The anomaly at the Bose-Einstein condensate is the first density anomaly observed in a quantum state. We propose that the mechanism underlying both the normal phase and the superfluid phase anomalies is related to zero point entropies and ground state phase transitions. A connection with the typical experimental scales and setups is also addressed. This key finding opens a new pathway for theoretical and experimental studies of water-like anomalies in the area of ultracold quantum gases.} 
 \end{abstract}


\maketitle

\indent The experimental realization of the Bose-Einstein condensation~\cite{Anderson1995ObservationVapor,Davis1995Bose-EinsteinAtoms} inaugurated a new era in physics by merging different areas, from condensed matter~\cite{Dutta2015Non-standardReview,Lewenstein2007UltracoldBeyond} to quantum information~\cite{Jaksch1999EntanglementCollisions,Cirac2004NewIons}. This landmark provided grounds for new applications involving the manipulation of ultracold atoms, from which  optical lattices, literal crystal arrays of light trapping neutral cold atoms~\cite{Greiner2002a,Bloch2005ExploringLattices}, stand as a prominent one. 
Among these applications, systems known as quantum simulators~\cite{Jaksch2005TheToolbox,Hofstetter2018QuantumSystems} have attained great importance since they can be used to experimentally implement and simulate scenarios for a plethora of theoretical ideas~\cite{Mueller2004ArtificialLiquids,Kapit2011Optical-latticeElectrodynamics}. Indeed, it is possible to  engineer them in highly controllable ways in regards to parameters such as dimensionality, lattice structure, composition and atomic interactions~\cite{Windpassinger2013EngineeringLattices}. In a theoretical level, the Bose-Hubbard model can be considered as a true prototype system, currently used to investigate quantum phase transitions, quantum coherence, and quantum computation~\cite{Fisher1989BosonTransition,Jaksch1998a,Sachdev2011QuantumTransitions}.\\
\indent  In this Letter we theoretically show that the density of bosons in optical lattices, described by the Bose-Hubbard model, anomalously increases with temperature at fixed pressure in both superfluid and normal fluid regimes. Such counter intuitive behavior, usually denominated as {\it density anomaly}, according to our analysis occurs at temperatures below $2.1$ nK (superfluid) and $17.7$ nK (normal fluid) for rubidium-87 atoms trapped in a simple cubic optical lattice. These anomalies are similar to those presented by liquid water between $0$ and $4^o$C at 1 atm~\cite{Kell1967PreciseAtmosphere,Debenedetti2003SupercooledWater} and, are useful to test the concept that thermodynamic waterlike anomalies arise from the competition between two scales of interaction, and appears associated with critical phenomena.\\
\indent An explanation for the thermodynamic and dynamic anomalous behavior of liquid water has been disputed through different thermodynamic scenarios. In the second critical point (SCP) hypothesis, which is based on computer simulations of the ST2 atomically detailed model of water~\cite{Poole1992PhaseWater}, followed by extensive investigations on other models for water~\cite{Gallo2016Water:Liquids},  the apparent divergence of thermodynamic response functions in a metastable region is consequence of a metastable liquid-liquid phase transition ending in a critical point~\cite{Poole1992PhaseWater,Stanley2008LiquidWater}. Nevertheless, this behavior in the case of water was never observed experimentally. The liquid-liquid transitions were  reported in models for carbon~\cite{Glosli1999Liquid-LiquidCarbon}, silicon~\cite{Sastry2003LiquidliquidSilicon}, silica~\cite{Saika-Voivod2005SimulatedSilica}, and experimentally observed in phosphorus~\cite{Monaco2003NaturePressure}, triphenyl phosphite, and n-butanol~\cite{Kurita2005OnSystems}. More recently, experiments with mixtures of water and glycerol~\cite{Murata2013GeneralSolutions} and measurements of correlations functions using time-resolved optical Kerr effect (OKE) of supercooled water~\cite{Taschin2013EvidenceConditions} favor the SCP hypothesis, despite debates in literature~\cite{Limmer2013TheII,Gallo2016Water:Liquids}.\\
\indent The suggested connection between thermodynamic anomalies and criticality is difficult to be tested experimentally since the system freezes before reaching the critical temperature as in the case of water or the thermodynamic anomalies are not clear. In addition, the complexity of the water structure makes difficult to unveil the connection between the microscopic interactions, thermodynamic anomalies and criticality. Due to its experimental manageability and for being numerically treatable, we propose to use the Bose-Hubbard model as a platform to establish this connection.\\
\begin{figure*}
\centering
\includegraphics[clip,scale=0.6]{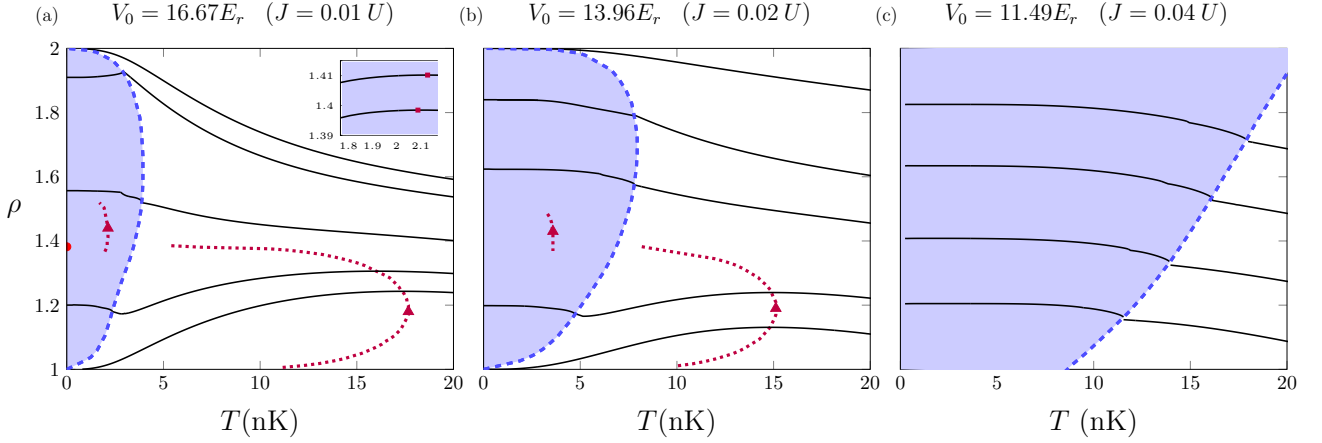}
\caption{The density $\rho$ as a function of temperature $T$ (in nano-Kelvin units) at fixed pressures for hopping amplitudes: (a) $J = 0.01\;U$, (b) $J = 0.02\;U$, and (c) $J = 0.04\;U$. The superfluid phase is highlighted in blue. Blue dashed lines denote the boundaries between superfluid and normal phases, while  purple dotted lines represents the TMD curves. The inset in (a) exhibit a zoom of isobaric curves in the superfluid phase.}
\label{fig1}
\end{figure*}
\indent  The dynamics of itinerant bosons in a lattice is governed by the Bose-Hubbard Hamiltonian
\begin{eqnarray}
H = - J \sum_{\langle i,j \rangle}{b}^{\dagger}_i{b}_j + \frac{U}{2} \sum_{i} n_i(n_i-1)- \mu \sum_{i} n_i \;,
\label{bose_hubbard_hamiltonian}
\end{eqnarray}
where $b^{\dagger}_i$, $b_i$, $n_i$ designates the bosonic creation, annihilation and number operators at site $i$, respectively; $\mu$ is the chemical potential. The parameter $U$ represents the on site interaction (typically repulsive, taking positive values) and $J$ accounts for the hopping amplitude, a kinetic term involving the probability of tunneling between first neighbor sites. \\
\indent In order to map the thermodynamics of the bosons we employ a variational and non-perturbative self-consistent approach, the Self-Energy Functional Theory~\cite{Hugel2016BosonicTheory}. The formalism, which comprehends previous BDMFT approaches~\cite{Hu2009DynamicalModel,Anders2011DynamicalBosons}, is based on a Legendre transform of the bosonic Baym-Kadanoff functional $\Gamma_{BK}=\Gamma_{BK}[\mathbf{\Phi},\mathbf{G}]$~\cite{Baym1961ConservationFunctions,Baym1962Self-ConsistentSystems,DeDominicis1964StationaryFormulation,DeDominicis1964StationaryFormulationb} from the one- and two-point propagators $\mathbf{\Phi}$ and $\mathbf{G}$ to their respective  self-energies $\mathbf{\Sigma_{1/2}}$ and $\mathbf{\Sigma}$~\cite{Potthoff2003Self-energy-functionalElectrons,Stefanucci2013NonequilibriumSystems}, with bold symbols denoting Nambu objects. The self-energy effective action obtained reads as $\Gamma_{SE} [\mathbf{\Sigma_{1/2}},\mathbf{\Sigma}] = \frac{1}{2}(\mathbf{F}-\mathbf{\Sigma_{1/2}})^{\dagger}\mathbf{G_0}(\mathbf{F}-\mathbf{\Sigma_{1/2}})+\frac{1}{2}\Tr \ln [-(\mathbf{G_0}^{-1}-\mathbf{\Sigma})] + \mathcal{F}[\mathbf{\Sigma_{1/2}},\mathbf{\Sigma}]$  where the universal functional $\mathcal{F}[\mathbf{\Sigma_{1/2}},\mathbf{\Sigma}]$ is the Legendre transform of the universal Luttinger-Ward functional $\Phi_{LW}[\mathbf{\Phi},\mathbf{G}]$, containing all two-particle irreducible diagrams. At the physical solution, $\Gamma_{SE}$ is stationary and equal to the free energy $\Omega = \Gamma_{SE}$~\footnote{As a result of $\Omega$ and $\Gamma_{SE}$ being connected by successive Legendre Transforms.}, yielding the Dyson's equations $\partial_{\mathbf{\Sigma_{1/2}}^{\dagger}} \Gamma_{SE} = -\mathbf{G_0}(\mathbf{F}-\mathbf{\Sigma_{1/2}})+\mathbf{\Phi}=0$ and $2 \partial_{\mathbf{\Sigma}} \:\Gamma_{SE} = -(\mathbf{G_0}^{-1}-\mathbf{\Sigma})^{-1}+\mathbf{G}=0$. The universality of the functional $\mathcal{F}$ enables us to overcome its complexity with the introduction of an exactly solvable reference system (denoted by primed quantities) exhibiting the same interactions as the original one. The functional evaluated at $\mathbf{\Sigma'_{1/2}}=\mathbf{\Sigma_{1/2}}$ and $\mathbf{\Sigma'}=\mathbf{\Sigma}$ can be expressed as
\begin{eqnarray}
\Gamma_{SE} [\mathbf{\Sigma'_{1/2}},\mathbf{\Sigma'}] = \Omega'+ \frac{1}{2}(\mathbf{F}-\mathbf{\Sigma'_{1/2}})^{\dagger}\mathbf{G_0}(\mathbf{F}-\mathbf{\Sigma'_{1/2}})+ \nonumber \\
-\frac{1}{2}(\mathbf{F'}-\mathbf{\Sigma'_{1/2}})^{\dagger}\mathbf{G'_0}(\mathbf{F'}-\mathbf{\Sigma'_{1/2}})+\frac{1}{2}\Tr \ln \left[\frac{\mathbf{G_0}^{-1}-\mathbf{\Sigma'}}{\mathbf{G'_0}^{-1}-\mathbf{\Sigma'}}\right]\;. \nonumber
\label{self_energy_functional}
\end{eqnarray}
The solution of the reference system provides a parametrization of the self-energies in terms of $\mathbf{F'}$ and $\mathbf{G'_0}$, according to $\Gamma_{SFT}[\mathbf{F'},\mathbf{G'_0}]=\Gamma_{SE} [\mathbf{\Sigma'_{1/2}}[\mathbf{F'},\mathbf{G'_0}],\mathbf{\Sigma'}[\mathbf{F'},\mathbf{G'_0}]] $. The approximation consists in constraining the variational principle to the subspace of self-energies of the reference system; this procedure yields the Euler equations $\partial_{\mathbf{F'}} \Gamma_{SFT} =0$ and $\partial_{\mathbf{G'_0}^{-1}} \Gamma_{SFT} =0$. In particular, we choose a local reference system, the SFA3 minimal construction~\cite{Hugel2016BosonicTheory}, comprehending three variational parameters: the $U(1)$ symmetry-breaking linear field $F'$ conjugated to the creation/annihilation operators ($b^{\dagger}$ and$b$), and the two fields, $\Delta_{00}$ and $\Delta_{01}$, coupled with the density ($b^{\dagger}b$) and pair creation/annihilation operators ($b^{\dagger}b^{\dagger}$ and $bb$), respectively. The Hamiltonian describing the bosonic state is given by
\begin{eqnarray}
H'[\mathbf{F'},\mathbf{\Delta}] = \frac{U}{2}n(n-1)- \mu n + \frac{1}{2}\mathbf{b}^{\dagger}\mathbf{\Delta}\mathbf{b}+\mathbf{F}'^{\dagger}\mathbf{b} \;,
\label{Hamiltonian_ref}
\end{eqnarray}
where $\mathbf{b}=(b,\; b^{\dagger})$, $\mathbf{F'}=(F',\; F'^*)$ and $\mathbf{\Delta} = \Delta_{00}\mathbf{1} + \Delta_{01}\sigma_{x}$. The numerical analysis concerning the stationary solutions $\nabla \Gamma_{SFT}[F',\Delta_{00},\Delta_{01}]=0$ and their agreement with the results exposed in reference~\cite{Hugel2016BosonicTheory} are detailed in the Supplemental Material.\\
\indent The behavior of the total density $\rho=-\frac{1}{V}\left(\frac{\partial \Omega}{\partial \mu}\right)_{T}$ with temperature $T$, volume $V$ and at fixed pressure $P$ is determined by the isobaric thermal expansion coefficient $\alpha =-\frac{1}{\rho}\left(\frac{\partial \rho}{\partial T}\right)_{P}$. For $\alpha<0$ density increases with temperature and a region of anomalous density behavior is identified by a temperature of maximum density line defined as $\alpha=0$. The thermal expansion at constant pressure $\alpha$ and at fixed chemical potential $\alpha_{\mu}$ are related by a simple change of variables $\alpha = \alpha_{\mu} -\frac{1}{\rho}\left(\frac{\partial \rho}{\partial \mu}\right)_{T}\left(\frac{\partial \mu}{\partial T}\right)_{P}\;$. As $T\rightarrow 0$, both $\left(\frac{\partial \rho}{\partial \mu}\right)_{T} \rightarrow 0$ and $\left(\frac{\partial \mu}{\partial T}\right)_{P} = \frac{s}{\rho}\rightarrow 0$, where the $s=-\frac{1}{V}\left(\frac{\partial \Omega}{\partial T}\right)_{\mu}$ is the entropy per volume. Hence, the coefficients $\alpha$ and $\alpha_{\mu}$ are interchangeable to investigate waterlike anomalies near the ground state~\cite{Rizzatti2018a}. The pressure is fixed employing the Gibbs-Duhem relation $dP = \rho d\mu + s dT = 0 $,  where $P$ is related to the grand-canonical potential according to $-PV= \Omega = \Gamma_{SFT}$. \\
\begin{table*}
\caption{Experimental parameters regarding potential depths $V_0$, scattering length $a_s$, and laser wavelengths $\lambda$ of optical lattices implemented using different alkali metal elements, for the hopping amplitudes $J=0.01\;U$ and $J=0.02\;U$. The maximum temperatures in which density anomalies are observed in superfluid and normal phases (the highlighted triangular points in Fig.~\ref{fig1}) are also addressed.}
\begin{tabular}{ c | c c | c c c | c c c }
\hline \hline
\multirow{2}{*}{Element} &             &               & \multicolumn{3}{c|}{$J=0.01\;U$}
& \multicolumn{3}{c}{$J=0.02\;U$}   \\
\cline{4-9}
                            & $\lambda$ (nm)& $a_s$ ($a_0$) &  $V_0/E_r$    & $T_{NA}$    (nK)    & $T_{SA}$    (nK)    &  $V_0/E_r$    & $T_{NA}$ (nK)   & $T_{SA}$    (nK)  \\
\hline
${}^{11}$Na~\cite{Jaksch1998a}          & 985  & 85            &  18.07        & 37.91                 & 4.57                & 15.25           & 32.61              & 7.73 \\
${}^{87}$Rb~\cite{Greiner2002a}           & 852  & 103           &  16.67        & 17.67                 & 2.13                & 13.96           & 15.12              & 3.59 \\
${}^{133}$Cs~\cite{Gemelke2009}          & 1064          & 460           &  11.87        & 20.54                 & 2.47                & 9.60           & 17.12              & 4.06 \\
\hline \hline
\end{tabular}
\label{table}
\end{table*}
\indent In order to connect the Bose-Hubbard model to an optical lattice experimental set up we use a prescription due to Zwerger~\cite{Zwerger2003MottLattices,Bloch2005ExploringLattices}, valid on the strong coupling field regime. In this approximation tunneling and on site interaction are related to the potential depth $V_0$ of a laser beam with wavelength $\lambda$ by the equations
\begin{eqnarray}
U \approx 4\sqrt{2\pi} \frac{a_s}{\lambda}\left(\frac{V_0}{E_r}\right)^{3/4} E_r,
\label{U}
\end{eqnarray}
and
\begin{eqnarray}
J \approx \frac{4}{\sqrt{\pi}} \left(\frac{V_0}{E_r}\right)^{3/4} e^{-2\sqrt{\frac{V_0}{E_r}}} E_r \;,
\label{J}
\end{eqnarray}
in units of the recoil energy $E_r=\frac{\hbar^2k^2}{2m}$, where $k$ is the corresponding lattice wavenumber, $m$ the atomic mass, and $a_s$ the scattering length of the $s$ wave of the trapped atom. The potential depth $V_0$, for a given ratio between hopping and on site interaction $J/U$, can be calculated from Eq.~(\ref{U}) and~(\ref{J}) with data from experimental setups for different alkali elements~\cite{Jaksch1998a, Greiner2002a, Gemelke2009}, as listed in Table~\ref{table}. It is thus possible to consider our theoretical results within the context of an specific optical trap implementation, from which we select a gas of rubidium-87~\footnote{Rubidium$-87$ was chosen since it is most common element used in optical lattices} in simple cubic optical lattice trapped to standing waves from pairs of lasers of $985$ nm wavelength~\cite{Greiner2002a}. \\
\indent In Figures~\ref{fig1}(a)-(c) illustrate the average number of $^{87}\mathrm{Rb}$ atoms per site versus temperature at fixed pressures as solid black lines for increasing hopping amplitudes or, equivalently, decreasing potential depths from (a) to (c). The superfluid to normal phase boundary is illustrated as a reentrant dashed blue line and the blue filled area represents the superfluid phase. Figures~\ref{fig1}(a)-(b) show that at high values of $V_0$ (low values of $J/U$) there are two regions in which density presents a local maximum, the Temperature of Maximum Density, TMD: one at the normal phase (NA) and another at the superfluid phase (SA).\\
\indent Figure~\ref{fig1}(a) portrays a large region in the density versus temperature phase diagrams where the NA is present. However, as the hopping increases as shown in Figure~\ref{fig1}(b) this region  occupies a smaller region in temperatures. In addition to the normal phase TMD, the superfluid phase also exhibits a density anomalous behavior illustrated in Fig.~\ref{fig1}(a), with a few superfluid isobaric densities drawn in the inset.  As the hopping becomes larger it dominates the free energy, leading the superfluid to occupy a larger region in the phase diagram and suppressing both superfluid and normal anomalies, as presented Fig.~\ref{fig1}(c). Although the phase diagram in Fig.~\ref{fig1} illustrates the case of rubidium-87, it can be adapted to other atoms by re-scaling temperatures through Eqs.~(\ref{U}) and (\ref{J}) with data from Table~\ref{table}.  This procedure was used to calculate the temperatures that must be achieved for experimentally detecting SA and NA for ${}^{11}$Na, ${}^{87}$Rb, and ${}^{133}$Cs, as listed in Table~\ref{table}~\footnote{Data for ${}^{11}$Na and ${}^{87}$Rb were extracted from experimental conditions of cubic lattice implementations, while data for ${}^{133}$Cs from a 2D experimental setup.}. In Fig.~\ref{fig1} (a) and (b), for rubidium-87, these points are marked as the triangular symbols over the TMD curves. \\
\indent Density anomalies in the normal fluid can be traced back to the ground state phase transitions between Mott Insulators of successive occupation numbers~\cite{Rizzatti2018a}. This anomalous behavior, present even in the absence of hopping, arises from the competition between the chemical potential, which promotes the boson occupation in the lattice, with the on site repulsion interaction $U$, which favors the boson removal. As the temperature increases entropy first favors filling up the sites but, for high enough temperatures, entropy increases by removing particles from the system to increase the mobility of the particles left. This is a classical behavior similar to that of liquid water, where bonding and non-bonding structures compete: at lower temperatures density increases by disrupting hydrogen bonds while at higher temperatures enhanced particles' velocities increase the available volume, decreasing density.  The novelty here is that this phase is not completely destroyed by the hopping, persisting for values of the J possible to be observed experimentally.\\
\indent The hopping, however, brings a new phenomena not observed for $J=0$, the SA, a quantum density anomaly. The physical origin of this behavior is also the competition between chemical potential and the repulsion $U$. But for the SA the TMD line appears at lower temperatures and higher densities when compared with the NA, because in this case the hopping adds to temperature. \\
\begin{figure}
\centering
\includegraphics[clip,scale=0.55]{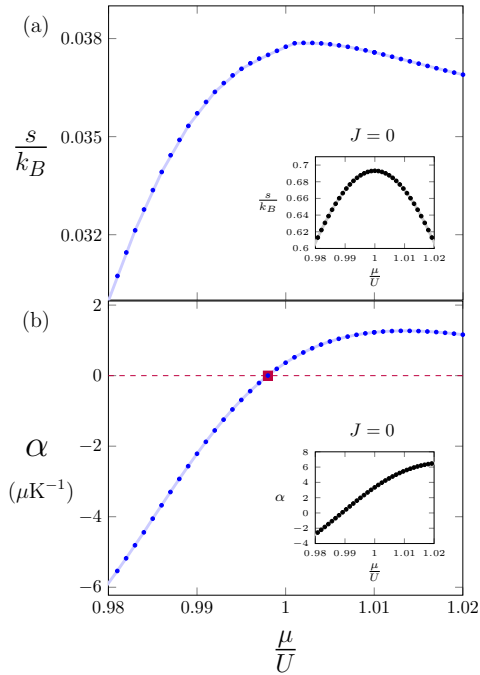}
\caption{The entropy (a) and thermal expansion (b) coefficient are exhibited as functions of the chemical potential $\mu$ for $V_0=16.67 E_r$ ($J=0.01\;U$) at $T=1.85$ nK ($k_BT=0.023\;U$) deep in the superfluid regime. The respective insets depict the atomic limit scenario.}
\label{fig2}
\end{figure}
\indent A physical insight into the mechanism behind the density anomaly can be extracted from the atomic limit. Inhibiting the hopping, a ground state degeneracy, related to a phase transition in number occupation between Mott Insulators, is settled  whenever the chemical potential $\mu$ reaches an integer value of the interaction $U$. At such transition points two states are equally accessible and this degeneracy accounts for an observed macroscopic residual entropy of $s_0=k_B\ln 2$. For finite temperatures, zero point entropies produce peaks near those points as the chemical potential is varied~\cite{Rizzatti2018a}. \\
\indent By turning on the tunneling probability adiabatically the superfluid phase emerges exactly from Mott Insulator transition points, mitigating residual entropies, as should be expected through the third law of thermodynamics. Indeed, a similar physical mechanism was observed on spin ice~\cite{Kato2015}, as the emergence a quantum spin ice phase with its delocalized spin states prevents the existence of a degenerate state given by ice rules~\cite{Pauling1935}. Thus, by turning on the hopping transition the previously mentioned entropy peak remain deep in the superfluid phase but is less prominent, as shown in figure~\ref{fig2} (a) for $V_0=16.67 E_r$ (or $J=0.01\;U$) and $T=1.85$ nK. Formally, the entropy peaks mark a change in the behavior of density with temperature according to the Maxwell relation
\begin{eqnarray}
\left(\frac{\partial s}{\partial \mu}\right)_{T}=\left(\frac{\partial \rho}{\partial T}\right)_{\mu}= -\rho \alpha_\mu\;,
\label{maxwell_relation}
\end{eqnarray}
which results in the sign flip of thermal expansion in the superfluid phase~\ref{fig2} (b).\\
\indent Indeed, for the NA regions, the residual entropies act as generators of density anomaly. Although the introduction of the hopping tends damp the residual entropy, such \textit{singularity} is so strong that it propagates its effect through the parameter space making the density anomaly occur even in the case where tunneling is allowed.
The peculiar way in which that reduction is settled, preserving the central peak, generates the density anomaly in the superfluid state. More precisely, when the tunneling probability is perturbative but finite, the $U(1)$ group symmetry of the Hamiltonian can be spontaneously broken, lifting the ground state degeneracy regarding the number occupation (as can be also checked in Fig.~\ref{fig1}, the ground state density can assume any continuum value inside the superfluid phase). The superfluid phase, which is born from  this process, mitigates the previously observed residual entropy, and reduces the thermal response function (where the third law of thermodynamics takes place). Nevertheless, the effect of the atomic limit phase transition propagates through parameter space and is felt even at finite hopping amplitudes, as illustrated by the insets of Fig.~\ref{fig2}. \\
\indent In Fig.~\ref{fig3}, we provide a physical visualization of the reported phenomenon in the real space considering a harmonic trapping field and using a local density approximation (LDA). We restrict such analysis to the normal phase anomaly since the variations in density with temperature are more prominent. The harmonic confinement potential is given by $V_h(r)= \frac{1}{2}m\omega^2 r^2$, where $r = \sqrt{x^2+y^2+z^2}$ is the radial distance from the center of the trap and the associated oscillation frequency $\omega$ is fixed at $\frac{\omega}{2\pi} = 65$ Hz, following the experiments of Greiner~\cite{Greiner2002a}. The number of lattice sites, spaced by a distance $a=\frac{\lambda}{2}$, is 60 for each coordinate direction (of lenght $L$), and the lattice depth is held at $V_0=16.67 E_r$ ($J=0.01 U$). Furthermore, in the LDA framework, the chemical potential across the lattice takes the form $\mu(r) = \mu_0 - V_h(r)$, the total number of particles is kept constant $N\approx 3.6\times 10^4$ as well as the total pressure $P=\int_{-\infty}^{\mu_0} \rho d\mu = \int_{0}^{L/2} \rho (r) \frac{d\mu}{dr}dr= 0.85 \frac{U}{a^3}$.  The densities profiles in the radial direction are shown in Fig.\ref{fig3}(a) for two different temperatures $T=6$ nK and $T=10$ nK. The Fig.\ref{fig3}(b) makes explicit the density variations $\Delta \rho$ between the different temperature profiles across the $xy$ plane ($z=0$). As the chemical potential varies with distance, anomalous regions of $\Delta \rho > 0$ alternates with the regular ones with $\Delta \rho <0$, yielding a wedding cake pattern. Therefore, this peculiar behavior can be regarded as a signature of the density anomaly.  \\
\begin{figure}
\centering
\includegraphics[clip,scale=0.65]{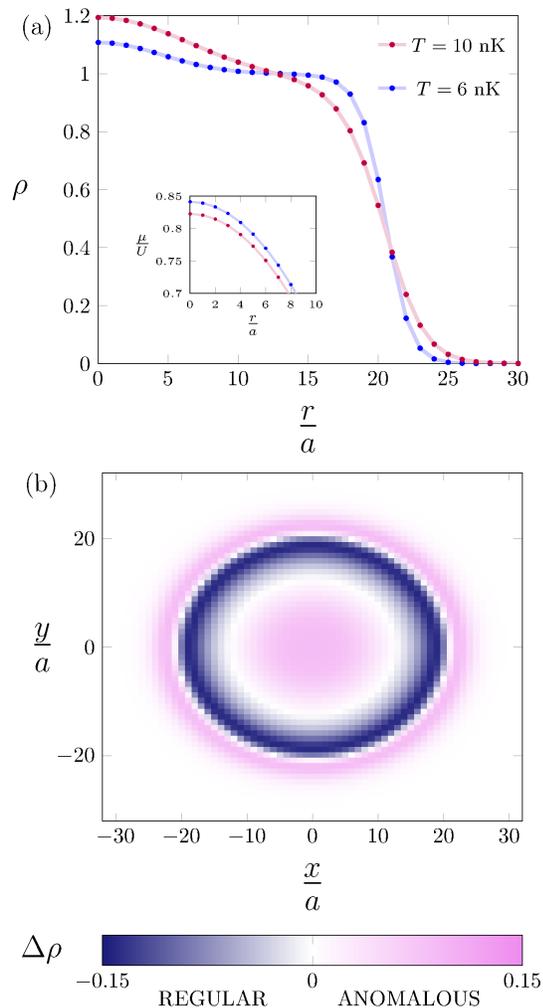}
\caption{(a) The radial density profiles for $T=6$ nK and $T=10$ nK, considering an optical trap with $V_0=16.67 E_r$ ($J=0.01\;U$) in a LDA scenario, with the respective parabolic chemical potentials portrayed as an inset. (b) The density variations $\Delta \rho$ related to the previous radial profiles are represented in the $z=0$ plane.}
\label{fig3}
\end{figure}
\indent In conclusion, we have predicted theoretically for the first time the occurrence of density anomaly in a quantum system considering parameters compatible with its experimental realization in optical lattices, within the framework described by the Self-Energy Functional Theory. It was also shown that the physical mechanism underlying normal density anomalies relies on the presence of a zero point entropy in the atomic limit, marking phase transitions between Mott Insulators with different occupation. The inclusion of the hopping amplitude (enabling the rise of a superfluid phase) lifts the ground state degeneracy, generate correlations among different sites and damps residual entropies and thermal expansion. Nevertheless, regions of anomalous density behavior can be found in a perturbative regime ($J \ll U$) corresponding to atomic recoil energy being much smaller than the intensities of the confining field $E_r \ll V_0 $. For very intense confining fields waterlike anomalies are also found inside the superfluid regime, as was illustrated for the case ${}^{87}$Rb in Fig.~\ref{fig1}. Our proposition is that by understanding the competition between different physical mechanisms contributing to free energy, usually manifested through interactions between particles (but here including chemical potential and hopping), and the relation between residual entropy and ground state phase transitions, it is possible to design and predict the phenomenology of density anomaly in systems other than liquid water, as illustrated here with optical lattices of rubidium-87, sodium-11 and cesium-133 atoms.


%

\end{document}